\documentclass[a4paper,12pt]{article}
\usepackage{amsmath}
\usepackage{amssymb}
\usepackage{amsfonts}
\usepackage{a4wide}
\usepackage{indentfirst}
\usepackage{times}
\usepackage{txfonts}

\newcommand {\oks}[2]{{\raise0.7ex\hbox{${\scriptstyle #1}$}\!\mathord{\left/
{\vphantom{{1}{2}}}\right.\kern-\nulldelimiterspace}\!\lower0.7ex
\hbox{${\scriptstyle #2}$}}}

\begin{document}

\title{\bf Neutrino-antineutrino pair production \\ by a photon in a dense matter}

\author{\bf A. E. Lobanov\thanks{E-mail: lobanov@phys.msu.ru}}
\date{}
\maketitle
\begin{center}
{\em Moscow State University, Department of Theoretical physics,
$119992$  Moscow, Russia}
\end{center}

\begin{abstract}
The possibility of radiative effects that are due to interaction
of fermions with a dense matter is investigated.
Neutrino-antineutrino photo-production is studied. The rate of
this process is calculated in the Furry picture. It is
demonstrated that this effect does not disappear even if  the
medium refractive index is assumed to be equal to unity. The rate
obtained strongly depends on the polarization states of the
particles involved. This leads to evident spatial asymmetries,
which may have certain consequences observable in astrophysical
and cosmological studies.
\end{abstract}

A Dirac massive neutrino has non-trivial electromagnetic
properties. In particular, it possesses a non-zero magnetic moment
\cite{FujShr80}. Therefore a Dirac massive neutrino propagating in
a dense matter can emit electromagnetic radiation due to the weak
interaction with background fermions \cite{L49,L55}. The
phenomenon was called the neutrino spin light in analogy with the
effect, related with the synchrotron radiation power depending on
the electron spin orientation (see \cite{BTB95}). The properties
of spin light were investigated in the quasi-classical approach
\cite{L49,L55,ST0410297}, and basing upon the consistent quantum
theory \cite{L57,L58,GST0502231}.

However this is not the only effect caused by the non-trivial
electromagnetic properties of neutrino. In the present work we
investigate the possibility of $\nu\bar{\nu}$ pair production by a
photon. As is evident from the following account, such process in
a dense matter is available even if the refractive index of the
medium is assumed to be equal to unity, i.e., the photon
dispersion law is the vacuum one.

It is well known  that the process  $\gamma \rightarrow \nu+
\bar{\nu}$ is kinematically possible whenever the photon four
momentum $k^{\mu}=\{k^{0},{\bf k}\}$ is time-like even if
neutrinos are assumed to be massless. This process, called
``plasmon decay'', can really take place in stars (see, e.g.,
\cite{Raffelt} and references therein). Unlike the ``plasmon
decay'', the reaction under investigation is due to the
modification of the neutrino dispersion law caused by the coherent
interaction of neutrino with a dense matter.

When the interaction of neutrino with the background fermions is
considered to be coherent, the propagation of neutrino in matter
is described by the Dirac equation with the effective potential
\cite{Wolf,MS}. In what follows, we restrict our consideration to
the case of a homogeneous and isotropic medium. Then in the
framework of the minimally extended standard model, the form of
this equation is uniquely determined by the assumptions similar to
those adopted in \cite{F}:
\begin{equation}\label{1}
  \left(i
  \hat{\partial}-\frac{1}{2}\hat{f}(1+\gamma^{5})-m_{\nu}\right)
  \varPsi_{\nu}=0.
\end{equation}
\noindent The function $f^{\mu} = {\mathrm{const}}$ is a linear
combination of fermion currents and polarizations. The quantities
with hats denote scalar products of the Dirac matrices with
4-vectors, i.e., $\hat{a} \equiv \gamma^{\mu}a_{\mu}.$

If the medium is at rest and unpolarized then ${\bf f}
=0.$ The component $f^{0}$ calculated in the first order of the
perturbation theory is as follows \cite{PP,N,NR}:
\begin{equation}\label{2}
f^{0}=\sqrt{2}G_{\mathrm F}\bigg\{\sum\limits_{f}^{}
\left(I_{e\nu} + T_{3}^{(f)}-2Q^{(f)}\sin^{2}\theta_{\mathrm
W}\right)(n_{f}-n_{\bar{f}})\bigg\}.
\end{equation}
\noindent Here, $n_{f},n_{\bar{f}}$ are the number densities of
background fermions and anti-fermions, $Q^{(f)}$ is the electric
charge of the fermion, and $T_{3}^{(f)}$ is the third component of
the weak isospin for the left-chiral projection of it. The
parameter \mbox{$I_{e\nu}$} is equal to unity for the interaction
of electron neutrino with electrons. In other cases $I_{e\nu}= 0.$
Summation is performed over all fermions of the background.

In what follows we use solutions of equation (\ref {1}) in which
quantum numbers of the problem are the kinetic momentum components
of neutrino, related to its group 4-velocity $u^{\mu}$ by the
relations $q^{\mu} = m_{\nu}u^{\mu},\; q^{2}= m^{2}_{\nu}.$ This
choice can be justified, since it is the particle kinetic momentum
that can be really observed.

When the medium is at rest and unpolarized then the
orthonormalized system of such solutions is\cite{L57,L58}:
\begin{equation}\label{21}
\varPsi(x)= \frac{\left|\Delta_{q \zeta}\right|}{\sqrt{2q^{0}}}
\,e^{-i(q^{0}+f^{0}/2)x^{0}} e^{i{\bf{qx}}\Delta_{q \zeta} }
(\hat{q}+m_{\nu})(1-\zeta\gamma^{5}\hat{S}_{tp} )\psi^0,
\end{equation}

\noindent where $\Delta_{q \zeta} = 1+\zeta f^{0}/2|{\bf{q}}|,$
and
\begin{equation}\label{0021}
S_{tp}^{\mu}=\frac{1}{m_{\nu}}\left\{|{\bf q}|,
 q^{0}{\bf q}/|{\bf q}|\right\},
\end{equation}
\noindent i. e., the eigenvalues $\zeta = \pm 1$ determine the
helicity of the particle. Here  $\psi^0$ is an arbitrary
normalized constant bispinor.

It is clear that the relation of the canonical momentum $\bf{P}$
to the kinetic momentum $\bf{q}$ is determined by the formula
\begin{equation}\label{03}
    {\bf P} = {\bf q}\Delta_{q \zeta},
\end{equation}
and the particle energy is given by the relation
\begin{equation}\label{003}
    \varepsilon = q^{0} +f^{0}/2 = \sqrt{{\bf q}^{2}+m^{2}_{\nu}}
    +f^{0}/2.
\end{equation}
It is seen that if the energy is expressed in terms of the kinetic
momentum, then it does not depend on the particle helicity, while
the particle canonical momentum is a function of the helicity.

Consider the neutrino-antineutrino pair photo-production process.
The formula for the neutrino pair creation probability by  a
polarized photon with the vacuum dispersion law is
\begin{equation}\label{ar1}
\begin{array}{c}
  \displaystyle P=-\frac{1}{2k^{0}}\!\int\!\! d^{4}x\,d^{4}y\!\int
  \frac{d^{4}p\,d^{4}q}{(2\pi)^{6}}\,
  \delta(p^{2}\!-m^{2}_{\nu})\delta(q^{2}\!-m^{2}_{\nu})
  \\[8pt]
\displaystyle \times {\mathrm{Sp}}
\big\{\varGamma_{\mu}(x)\varrho_{\bar{\nu}}(x,y;p,\zeta_{\bar{\nu}})
\varGamma_{\nu}(y)\varrho_{\nu}(y,x;q,\zeta_{\nu})\big\}
\varrho^{\mu\nu}_{ph}(y,x;k).
\end{array}
\end{equation}
\noindent \noindent Here,
$\varrho_{\nu}(y,x;q,\zeta_{\nu}),\varrho_{\bar{\nu}}(x,y;p,\zeta_{\bar{\nu}})$
are the neutrino and antineutrino density matrices respectively,
$\,\varrho^{\mu\nu}_{ph}(y,x;k)$ is the initial photon density
matrix, $\varGamma^{\mu} = -\,\sqrt{4\pi}
\mu_{0}\sigma^{\mu\nu}k_{\nu}$ is the vertex function, where
$\mu_{0}$ is the anomalous magnetic moment of the neutrino. The
density matrices of longitudinally polarized antineutrino and
neutrino in the unpolarized matter at rest constructed with the
use of the solutions (\ref{21}) of equation (\ref{1}) have the
form
\begin{equation}\label{arx4}
\begin{array}{l}
\displaystyle \varrho_{\bar{\nu}}(x,y;p,\zeta_{\bar{\nu}})=
\frac{1}{2}\Delta_{p\zeta_{\bar{\nu}}}^{2}(\hat{p}- m_{\nu})
(1-\zeta_{\bar{\nu}}\gamma^{5}\hat{S}_{tp}(p))
e^{i(x^{0}-y^{0})(p^{0}-f^{0}/2)-i({\bf x}-{\bf y}) {\bf
p}\Delta_{p\zeta}},\\[10pt]
\displaystyle \varrho_{{\nu}}(y,x;q,\zeta_{\nu})=\frac{1}{2}
\Delta_{q\zeta_{\nu}}^{2}(\hat{q}+m_{\nu})
(1-\zeta_{\nu}\gamma^{5}\hat{S}_{
tp}(q))e^{i(x^{0}-y^{0})(q^{0}+f^{0}/2)-i({\bf x}-{\bf y}) {\bf
q}\Delta_{q\zeta_{\nu}}}.
\end{array}
\end{equation}

After integrating with respect to coordi\-nates we obtain the
expression for the transition rate under investi\-ga\-tion:
\begin{equation}\label{arxx4}
\begin{array}{c}
\displaystyle W= \frac{\mu^{2}_{0}}{k^{0}}\!\!\int
  \frac{d^{4}p\,d^{4}q}{(2\pi)}\,
  \delta(p^{2}\!-m^{2}_{\nu})\delta(q^{2}\!-m^{2}_{\nu})\delta(k^{0}-p^{0}-q^{0})
  \delta^{3}
  ({\bf
  k} -{\bf p}\Delta_{p\zeta_{\bar{\nu}}}-{\bf q}\Delta_{q\zeta_{\nu}})T(p,q),
\end{array}
\end{equation}
\noindent where
\begin{equation}\label{arxxx4}
\begin{array}{c}
T(p,q)=2\Delta_{p\zeta_{\bar{\nu}}}^{2}\Delta_{q\zeta_{{\nu}}}^{2}
\left\{(pk)(qk)+\zeta_{\bar{\nu}}\zeta_{{\nu}}\left[k^{0}|{\bf
p}|-p^{0}({\bf
  p}{\bf
  k})/|{\bf p}|\right]\left[k^{0}|{\bf q}|-q^{0}({\bf
  q}{\bf
  k})/|{\bf q}|\right]\right. \\[6pt]
  -\left. g\zeta_{\bar{\nu}}(qk)\left[k^{0}|{\bf p}|-p^{0}({\bf
  p}{\bf
  k})/|{\bf p}|\right]-g\zeta_{{\nu}}(pk)\left[k^{0}|{\bf q}|-q^{0}({\bf
  q}{\bf
  k})/|{\bf q}|\right]\right\}.
\end{array}
\end{equation}
\noindent Here $g= \pm 1$ denotes the sign of the initial photon
circular polarization.

After integrating over angular variables of antineutrino and
$|{\bf p}|,|{\bf q}|,$ we obtain the spectral-angular distribution
of the final neutrino
\begin{equation}\label{q3}
\begin{array}{c}
\displaystyle W=
\zeta_{\bar{\nu}}\zeta_{\nu}\frac{\mu^{2}_{0}}{4\pi{k^{0}}}
\!\!\int\!\frac{dp^{0}dq^{0}}{|{\bf p}||{\bf
q}|}\,\delta(k^{0}-p^{0}-q^{0})\,
{\mathrm{sign}}(\Delta_{p\zeta_{\bar{\nu}}} \Delta_{q\zeta_{\nu}})
\\[12pt]
\displaystyle\times\int\!\! dO\:\delta\left(\sqrt{ |{\bf
p}|^{2}\Delta_{p\zeta_{\bar{\nu}}}^{2}- (k^{0}-|{\bf
q}|\Delta_{q\zeta_{\nu}}\cos\vartheta_{\nu})^{2}} - |{\bf
q}||\Delta_{q\zeta_{\nu}}|\sin\vartheta_{\nu}\right)\\[12pt]
\displaystyle \!\!\!\!\times
\left[(f^{0}/2)(\zeta_{\bar{\nu}}q^{0}|{\bf
p}|+\zeta_{{\nu}}p^{0}|{\bf q}|)- k^{0}m^{2}_{\nu} -
g(f^{0}/2)\left(\zeta_{\bar{\nu}}\zeta_{{\nu}}|{\bf p}||{\bf
q}|+m^{2}_{\nu}+p^{0}q^{0}\right)\right]^{2}\!,
\end{array}
\end{equation}
\noindent where $$|{\bf p}|=\sqrt{(p^{0})^{2}-m^{2}_{\nu}},\quad
|{\bf q}|=\sqrt{(q^{0})^{2}-m^{2}_{\nu}},$$  $\vartheta_{\nu}$ is
the angle between the direction of the neutrino propagation and
the photon wave vector, and $dO=\sin \vartheta_{\nu}
d\vartheta_{\nu} d\varphi_{\nu}$ is the solid angle element.

Spectral-angular distribution of the final antineutrino can be
found after substitution
\begin{equation}\label{qqq10}
q^{0}\leftrightarrow p^{0},\quad |{\bf q}|\leftrightarrow|{\bf
p}|,\quad \zeta_{{\nu}}\leftrightarrow\zeta_{\bar{\nu}},\quad
\vartheta_{\nu}\rightarrow\vartheta_{\bar{\nu}},
\end{equation}
made in equation (\ref{q3}). Naturally, the kinetic momenta of
neutrino ${\bf q}$ and antineutrino ${\bf p}$ are in a plane with
the photon momentum ${\bf k}.$ If
${\mathrm{sign}}(\Delta_{p\zeta_{\bar{\nu}}}
\Delta_{q\zeta_{\nu}}) =1$, the kinetic momenta ${\bf p},{\bf q}$
have opposite azimuthal angles. However if
${\mathrm{sign}}(\Delta_{p\zeta_{\bar{\nu}}}
\Delta_{q\zeta_{\nu}})=-1$, this vectors have the same azimuthal
angles. It should be emphasized that this fact is rather unusual.

It is convenient to express the results  of integrating over
angular variables of neutrino in terms of dimensionless
quantities. Introducing the notations
\begin{equation}\label{q10}
\begin{array}{lll}
\eta = k^{0}/2m_{\nu},\quad & p^{0}/m_{\nu}=\eta+y,\quad &
  q^{0}/m_{\nu}=\eta-y,\\
  d=|f^{0}|/2m_{\nu},\quad &
  \bar{\zeta}_{\bar{\nu},\nu}={\zeta}_{\bar{\nu},\nu}\,{\mathrm{sign}}(f^{0}),\quad
  &
  \bar{g}={g}\,{\mathrm{sign}}(f^{0}),
\end{array}
\end{equation}
we have
\begin{equation}\label{q16}
\begin{array}{c}
\displaystyle W_{\bar{g}\bar{\zeta}_{\bar{\nu}}\bar{\zeta}_{\nu}}
= \frac{\mu_{0}^{2}m_{\nu}^{3}}{8\eta^{2}}\int^{y_{+}}_{y_{-}}\!\!
  \frac{dy}{\sqrt{(\eta+y)^{2}-1}\sqrt{(\eta-y)^{2}-1}}\,\\[16pt]
 \displaystyle\times \left[
  d\left(\bar{\zeta}_{\bar{\nu}}(\eta-y)\sqrt{(\eta+y)^{2}-1}+\bar{\zeta}_{\nu}
  (\eta+y)\sqrt{(\eta-y)^{2}-1}\right)  -2\eta    \right. \\[10pt]
 \displaystyle -\left.\bar{g}d\left(\bar{\zeta}_{\bar{\nu}}\bar{\zeta}_{\nu}
\sqrt{(\eta+y)^{2}-1}\sqrt{(\eta-y)^{2}-1}+(\eta + y)(\eta - y) +1
  \right)\right]^{2}.
\end{array}
\end{equation}
\noindent For pair production the necessary condition on the
photon energy is $\eta \geqslant 1.$ In fact the threshold of the
reaction can lie higher. So the integration limits $y_ \pm $ in
formula (\ref{q16}) are different for different values of
parameter $d.$

When $d<\oks{1}{2}$ we have
\begin{equation}\label{q011}
\begin{array}{ll}
y\in\varnothing & \quad \eta \in [1,\eta_{1}), \\
y\in [-y_{0},y_{0}] & \quad \eta \in [\eta_{1},\infty),
\end{array}
\end{equation}
\noindent if $\bar{\zeta}_{\bar{\nu}}=1, \bar{\zeta}_{{\nu}}=1,$
\begin{equation}\label{q11}
\begin{array}{ll}
y\in \varnothing & \quad \eta \in [1,\infty),
\end{array}
\end{equation}
\noindent if $\bar{\zeta}_{\bar{\nu}}=-1, \bar{\zeta}_{{\nu}}=-1,$
\begin{equation}\label{q12}
\begin{array}{ll}
y\in \varnothing & \quad \eta \in [1,\infty),
\end{array}
\end{equation}
\noindent if $\bar{\zeta}_{\bar{\nu}}=-1, \bar{\zeta}_{{\nu}}=1,$
and
\begin{equation}\label{q012}
\begin{array}{ll}
y\in \varnothing & \quad \eta \in [1,\infty),
\end{array}
\end{equation}
\noindent if $\bar{\zeta}_{\bar{\nu}}=1, \bar{\zeta}_{{\nu}}=-1.$

When $\oks{1}{2}\leqslant d\leqslant{1}$ we have
\begin{equation}\label{q0111}
\begin{array}{ll}
y\in\varnothing & \quad \eta \in [1,\eta_{1}), \\
y\in [-y_{0}, y_{0}] & \quad \eta \in [\eta_{1},
\eta_{2}),\\
y\in [-(\eta-1),\eta-1] & \quad \eta \in [\eta_{2},\infty),
\end{array}
\end{equation}
\noindent if $\bar{\zeta}_{\bar{\nu}}=1, \bar{\zeta}_{{\nu}}=1,$
\begin{equation}\label{q111}
\begin{array}{ll}
y\in \varnothing & \quad \eta \in [1,\infty),
\end{array}
\end{equation}
\noindent if $\bar{\zeta}_{\bar{\nu}}=-1, \bar{\zeta}_{{\nu}}=-1,$
\begin{equation}\label{q121}
\begin{array}{ll}
y\in \varnothing & \quad \eta \in
[1,\eta_{2}),\\
y\in [-(\eta-1),-y_{0}] & \quad \eta \in
[\eta_{2},\infty),\\
\end{array}
\end{equation}
\noindent if $\bar{\zeta}_{\bar{\nu}}=-1, \bar{\zeta}_{{\nu}}=1,$
and
\begin{equation}\label{q0121}
\begin{array}{ll}
y\in \varnothing & \quad \eta \in
[1,\eta_{2}),\\
y\in [y_{0},\eta-1] & \quad \eta \in [\eta_{2},\infty),\\
\end{array}
\end{equation}
\noindent if $\bar{\zeta}_{\bar{\nu}}=1, \bar{\zeta}_{{\nu}}=-1.$

In the case of high matter density $(d>1)$ we have
\begin{equation}\label{q0110}
\begin{array}{ll}
y\in [-(\eta-1),\eta-1] & \quad \eta \in [1,\infty),
\end{array}
\end{equation}
\noindent if $\bar{\zeta}_{\bar{\nu}}=1, \bar{\zeta}_{{\nu}}=1,$
\begin{equation}\label{q110}
\begin{array}{ll}
y\in [-(\eta-1),\eta-1] & \quad \eta \in [1,\eta_{1}),\\
y\in [-(\eta-1),-y_{0}]\!\bigcup[y_{0},\eta-1] &
\quad \eta \in [\eta_{1},\eta_{2}),\\
y\in \varnothing &
\quad \eta \in [\eta_{2},\infty),\\
\end{array}
\end{equation}
\noindent if $\bar{\zeta}_{\bar{\nu}}=-1, \bar{\zeta}_{{\nu}}=-1,$
\begin{equation}\label{q120}
\begin{array}{ll}
y\in [-(\eta-1),\eta-1] & \quad \eta \in [1,\eta_{2}),\\
y\in [-(\eta-1),y_{0}] & \quad \eta \in
[\eta_{2},d),\\
y\in [-(\eta-1),-y_{0}] & \quad \eta \in
[d,\infty).\\
\end{array}
\end{equation}
\noindent if $\bar{\zeta}_{\bar{\nu}}=-1, \bar{\zeta}_{{\nu}}=1,$
and
\begin{equation}\label{q0120}
\begin{array}{ll}
y\in [-(\eta-1),\eta-1] & \quad \eta \in [1,\eta_{2}),\\
y\in [-y_{0},\eta-1] & \quad \eta \in
[\eta_{2},d),\\
y\in [y_{0},\eta-1] & \quad \eta \in [d,\infty).\\
\end{array}
\end{equation}
\noindent if $\bar{\zeta}_{\bar{\nu}}=1, \bar{\zeta}_{{\nu}}=-1.$

\noindent Here
\begin{equation}\label{q13}
\eta_{1}= \frac{1+d^{2}}{2d},\qquad \eta_{2}=
\frac{d^{2}}{2d-1},\qquad y_{0}=\frac{|\eta -d|\sqrt{\eta
-\eta_{1}}}{\sqrt{\eta -d/2}}.
\end{equation}

It is easy to verify that if parameter $\bar{\zeta}_{\nu}= -1$ for
the generated neutrino, then $\Delta_{q {\bar{\zeta}_{\nu}}} < 0.$
In a similar way,  if parameter $\bar{\zeta}_{\bar{\nu}}= -1$ for
generated antineutrino, then $\Delta_{p {\bar{\zeta}_{\bar{\nu}}}}
< 0.$ Therefore, such particles have  $p^{0} <
m_{\nu}\sqrt{1+d^{2}},$ or $q^{0} < m_{\nu}\sqrt{1+d^{2}},$ i.e.,
their energies lie lower than the spin light emission threshold
\cite{L57,L58}. Since the condition $\bar{\zeta} = -1$ is
necessary for photon emission by a neutrino in matter, the cascade
process of the form
$$
\gamma\rightarrow \nu + \bar{\nu} \rightarrow \nu + \bar{\nu} +
\gamma
$$
is impossible in our model.

The transition rate under investi\-ga\-tion is defined as
\begin{equation}\label{j0}
    W_{\bar{g}\bar{\zeta}_{\bar{\nu}}\bar{\zeta}_{\nu}}
    =\frac{\mu_{0}^{2}m^{3}_{\nu}}{12\eta^{2}}\left(J(y_{+})-J(y_{-})\right).
\end{equation}
\noindent Here
\begin{equation}\label{j1}
\begin{array}{c}
   \displaystyle J(y) = {d\eta}\left(4d\eta^{2}+d
    +6\bar{g}\eta\right)E(\chi,s)-{\eta}\left(2d^{2}-3
    \right)F(\chi,s) \\[8pt]
    \displaystyle +\frac{d}{2\eta}
    \left(2d(y^{2}-\eta^{2})-3d
    -6\bar{g}\eta\right)\sin\chi{\displaystyle
    \sqrt{\left((\eta-y)^{2}-1\right)\left((\eta+y)^{2}-1\right)}}
    \\[8pt]
    + \displaystyle \bar{\zeta}_{\bar{\nu}}\bar{\zeta}_{{\nu}}{d}y
    \left(2d(3\eta^{2}-y^{2}) +3d
    +6\bar{g}\eta\right) \\[8pt]
 \displaystyle - 2\bar{\zeta}_{{\nu}}{d}\left(3\eta -\bar{g}d(y^{2} -2\eta^{2}-y\eta
    -1)\right)\sqrt{\left((\eta+y)^{2}-1\right)}\\[8pt]
    + \displaystyle
    2\bar{\zeta}_{\bar{\nu}}d\left(3\eta -\bar{g}d(y^{2} -2\eta^{2}+y\eta
    -1)\right)\sqrt{\left((\eta-y)^{2}-1\right)},
\end{array}
\end{equation}
where $F(\chi,s),E(\chi,s)$ are elliptic integrals \cite{BE} of
the arguments
\begin{equation}\label{j3}
    \chi = \arcsin \frac{2y\eta }{y^{2}+\eta^{2}-1},\qquad
    s=\sqrt{1-\eta^{-2}}\,.
\end{equation}
If we consider the case with $d \ll 1, $ the following approximate
expression can be obtained
\begin{equation}\label{j5}
\begin{array}{c}
\displaystyle
W_{\bar{g}\bar{\zeta}_{\bar{\nu}}\bar{\zeta}_{\nu}}\approx
    \frac{\mu_{0}^{2}m_{\nu}^{3}}{16\eta}(1+\bar{\zeta}_{\bar{\nu}})
    (1+\bar{\zeta}_{{\nu}})\left\{
     \ln \frac{1+ \sqrt{1-\eta_{1}/\eta}}
    {1- \sqrt{1-\eta_{1}/\eta}}\,(1+
    \bar{g})\right.\\[12pt]\displaystyle
    +\left.\left[\frac{4}{3}\left(\frac{\eta}{\eta_{1}}\right)^{2}\!\!
    \sqrt{({1-\eta_{1}/\eta})^{3}}-2\sqrt{{1-\eta_{1}/\eta}}+
     \ln \frac{1+ \sqrt{1-\eta_{1}/\eta}}
    {1- \sqrt{1-\eta_{1}/\eta}}\right]\!(1-
    \bar{g})\!\right\}.
\end{array}
\end{equation}
\noindent Near the reaction threshold, defined by the relation
$\eta =
    \eta_{1}$, the transition rate is described
by the formula
\begin{equation}\label{jj3}
 W_{\bar{g}\bar{\zeta}_{\bar{\nu}}\bar{\zeta}_{\nu}}\approx
    \frac{\mu_{0}^{2}m_{\nu}^{3}}{8\eta}(1+\bar{\zeta}_{\bar{\nu}})
    (1+\bar{\zeta}_{{\nu}})\left[
    \sqrt{1-\eta_{1}/\eta}(1+
    \bar{g})+\sqrt{(1-\eta_{1}/\eta)^{3}}(1-
    \bar{g})\right],
\end{equation}
\noindent and for $d\eta \gg 1$, by the formula
\begin{equation}\label{j4}
W_{\bar{g}\bar{\zeta}_{\bar{\nu}}\bar{\zeta}_{\nu}}\approx
    \frac{\mu_{0}^{2}m_{\nu}^{3}}{3}d^{2}\eta(1+\bar{\zeta}_{\bar{\nu}})
    (1+\bar{\zeta}_{{\nu}})(1-
    \bar{g}).
\end{equation}
When $d\gtrsim 1, \eta \gg d$ the transition rate is described by
the formula (\ref{j4}) as well.

Pairs produced have zero angular orbital momentum if $\bar{g}=1$
(``allowed'' transition), and non-zero angular orbital momentum if
$\bar{g}=-1$ (``forbidden'' transition). Evidently, only high
energy photons with the helicity sign opposite to the sign of the
effective potential ($\bar{g}=-1$) can effectively produce
neutrino and antineutrino. For high energy photons with
$\bar{g}=1$ pair production is suppressed.

With the use of the effective potential calculated in the first
order of the perturbation theory (\ref{2}) the following
conclusions can be made. Let us discuss, as an example, the
neutron medium. The sign of the effective potential  is negative
in this case, so only right-handed polarized photons of high
energy can effectively interact with such medium. In the
ultra-relativistic limit, (here we use gaussian units), we have
for the rate of pair production
\begin{equation}\label{q29}
\displaystyle
W_{\bar{g}\bar{\zeta}_{\bar{\nu}}\bar{\zeta}_{\nu}}\approx
\frac{\alpha\varepsilon_{\gamma}}{192\,\hbar}
\left(\frac{\mu_{0}}{\mu_{\mathrm B}}\right)^{2}
\left(\frac{{G}_{{\mathrm F}}\,n_{n}}{m_{e}c^{2}}\right)^{2}
(1-\bar{\zeta}_{i})(1+\bar{\zeta}_{f})(1- \bar{g}).
\end{equation}
\noindent Here $\varepsilon_{\nu}$ is the neutrino energy,
$\mu_{\mathrm B}=e\hbar/2m_{e}c$ is the Bohr magneton, $\alpha $
is the fine structure constant, $m_{e}$ is the electron mass, and
${G}_{{\mathrm F}}$ is the Fermi constant. This formula is valid
for neutrinos of different flavors.

As a result of the reaction the left-handed polarized particles
essentially arise: an active left-handed polarized neutrino and a
practically ``sterile'' left-handed polarized antineutrino.
Electron neutrino can react with neutron medium as follows
\begin{equation*}
    \nu_{e} + n \rightarrow p^{+} + e^{-},
\end{equation*}
and $\nu_{\mu}, \nu_{\tau}$ cannot interact with the medium.

It is very interesting to consider the limit  $\,m_{\nu}
\rightarrow 0\,$ in eq. (\ref{q16}), when the neutrino anomalous
magnetic moment $\mu_{0}$ is supposed to be constant. We have
\begin{equation}\label{qqq16}
\begin{array}{c}
\displaystyle W_{\bar{g}\bar{\zeta}_{\bar{\nu}}\bar{\zeta}_{\nu}}
=
\frac{\mu_{0}^{2}k^{0}(f^{0})^{2}}{24}\left[(1+\bar{\zeta}_{\bar{\nu}})
    (1+\bar{\zeta}_{{\nu}})(1-
    \bar{g})+(1-\bar{\zeta}_{\bar{\nu}})
    (1-\bar{\zeta}_{{\nu}})(1+
    \bar{g})\,\Theta(\,|f^{0}|-2k^{0})\right].
\end{array}
\end{equation}
It is evident that the right-hand side of eq. (\ref{qqq16})
vanishes if we use the standard two-component model of neutrino
($\zeta_{\bar{\nu}}= 1, \zeta_{\nu} = -1$). Hence the rate of the
process is not equal to zero due to "sterile" states of neutrinos.

Recently \cite{ZLM} the possibility of electron-positron pair
production by a photon was investigated in the framework of
Standard Model Extension \cite{2,3,4,5} with axial-vector Lorentz
breaking background. In this model the Dirac equation for an
electron differ from (\ref{1}) only by the term that can be gauge
away. In the relativistic case, the formulas obtained in
\cite{ZLM} for the rate of pair production coincide with those for
``allowed'' transition ($\bar{g}=1$) in our model (see
(\ref{j5})). The transition rate decreases in the region of high
energies as it must be in the renormalizable theory. It is
significant that in our model based on the non-minimal interaction
and, therefore, non-renormalizable, ``forbidden'' transitions play
the main role at high energies.

In conclusion a few words about approximations used in our paper.
It is possible to suppose that the interaction of neutrino with
matter to be coherent if there are significant number of matter
particles in the de Broglie cell of the neutrino. This condition
leads to inequality
\begin{equation}\label{z1}
    \frac{n}{\gamma} \left(\frac{\hbar}{m_{\nu}c}\right)^{3}  \gg 1.
\end{equation}
If the matter number density is assumed to be  $n \approx
10^{38}{\mathrm{cm}}^{-3}$ which corresponds to the number density
of a neutron star, we have
\begin{equation}\label{z2}
\varepsilon_{\nu} \ll
10^{24}\left(\frac{m_{\nu}}{1{\mathrm{eV}}}\right)^{-2}{\mathrm{eV}}.
\end{equation}
Since the neutrino mass is estimated as $m_{\nu}\lesssim 1$eV the
quantity in the right-hand side of this inequality is close to the
Plank energy.

In our calculations we used the approach based on the static value
for the anomalous magnetic moment of neutrino. Unfortunately, the
conditions of applicability of this approach at high energies are
not known. However we believe that the results obtained in the
present paper may provide some hints to possible future
observation of the effects that are due to coherent neutrino
interaction with a dense matter. The special feature of the
results obtained in the present paper is that the pair production
rate strongly depends on the helicity of the particles, and this
leads to evident spatial asymmetries in the reaction products,
which may have certain consequences observable in astrophysical
and cosmological studies.

\bigskip

The author is grateful to A. V.~Borisov,  A. E.~Shabad, and V.
Ch.~Zhukovsky for fruitful discussions.

\bigskip

This work was supported in part by the grant of President of
Russian Federation for leading scientific schools (Grant SS ---
5332.2006.2).

\end{document}